\documentclass{sig-alternate}

\usepackage[usenames,dvipsnames]{color}
\usepackage{tabularx}
\usepackage{subfig}
\usepackage{multirow}
\usepackage{booktabs}
\usepackage{threeparttable}
\usepackage{verbatim}
\usepackage{ifthen}
\usepackage{mdwlist}
\usepackage{xspace}
\usepackage{balance}  
\usepackage{graphics} 
\usepackage{url}      

\newboolean{VIEWNOTES}
\setboolean{VIEWNOTES}{false}
\newboolean{ANONYMOUS}
\setboolean{ANONYMOUS}{false}

\newcommand{\ie}{\textit{i.e.}\xspace}
\newcommand{\eg}{\textit{e.g.}\xspace}

\newcommand{\etal}{\textit{et al.}\xspace}

\newcommand{\superscript}[1]{\ensuremath{^{\textrm{#1}}}}

\newcommand{\draftnote}[1]{{\ifthenelse {\boolean{VIEWNOTES}}
{\textcolor{ForestGreen}{NOTE: #1}}{}}}
\newcommand{\draftfnote}[1]{{\ifthenelse {\boolean{VIEWNOTES}}
{\footnote{\textcolor{ForestGreen}{#1}}}{}}}

\newcommand{\anon}[2][*****]{{\ifthenelse {\boolean{ANONYMOUS}}
{\textcolor{Black}{#1}}{#2}}}

\newcommand{\marginnote}[1]{{\ifthenelse {\boolean{VIEWNOTES}}
{\marginpar{\textcolor{ForestGreen}{\small{#1}}}}{}}}

\newcommand{\vmove}{\textit{MS}\xspace}

\newcommand{\tstill}{\textit{PL}\xspace}
\newcommand{\rmove}{\textit{MR}\xspace}
\newcommand{\rclick}{\textit{CR}\xspace}

\newcommand{\pstill}{\textit{PTS}\xspace}

\newcommand{\smt}{smt\superscript{2}\xspace}

\makeatletter
\def\url@leostyle{%
  \@ifundefined{selectfont}{\def\UrlFont{\sf}}{\def\UrlFont{\small\bf\ttfamily}}}
\makeatother
\begin{document}



\title{An Exploration of Cursor tracking Data}

\numberofauthors{2}
\author{
  \alignauthor {David Warnock}\titlenote{This work was carried out while undertaking an internship at Yahoo Labs in Barcelona.}\\
    \affaddr{\anon{Kelvin Connect Ltd}}\\
    \email{\anon{dave@davewarnock.com}}
  \alignauthor {Mounia Lalmas}\\
    \affaddr{{Yahoo Labs}}\\
    \email{{mounia@acm.com}}\\    
}


\maketitle

%
%
%
%
%
%


\begin{abstract}
Cursor tracking data contains information about website visitors which may provide new ways to understand visitors and their needs. This paper presents an Amazon Mechanical Turk study where participants were tracked as they used modified variants of the Wikipedia and BBC News websites. Participants were asked to complete reading and information-finding tasks. The results showed that it was possible to differentiate between users reading content and users looking for information based on cursor data. The effects of website aesthetics, user interest and cursor hardware were also analysed which showed it was possible to identify hardware from cursor data, but no relationship between cursor data and engagement was found. The implications of these results for web analytics and the design of user engagement experiments are discussed.
\end{abstract}






\section{Introduction}\label{sec_introduction}

Internet-based businesses rely on a variety of measurements to evaluate performance, improve services, target advertising and personalise a website for its visitors. Analysis is generally based on data such as page views, browsing history and search terms. However, it is likely that more data can be gathered by examining interactions such as cursor movements. Cursor tracking involves recording the position of the cursor as a user interacts with a webpage. Only a small number of existing analytic tools offer cursor tracking, and those that do generally provide heatmaps of the data showing the most common cursor positions. While this could be useful when testing usability, it does not provide data that can be used for user profiling.\\

Many researchers have argued that cursor tracking data can provide a new way to learn about website visitors \cite{Edmonds2007,Huang2012}. Existing work shows that cursor movements often correlate with gaze \cite{Chen2001,Huang2012,Rodden2008}, which suggests that some of the techniques employed in gaze tracking studies could have analogues suitable for use with cursor tracking data. If so, this could allow inexpensive and large-scale usability testing to be carried out `in the wild' using analysis methods that could previously only be used in a lab-based setting. Other researchers have suggested that cursor tracking data could reveal user engagement \cite{Edmonds2007}, age and disability \cite{Riviere1996,Trewin1999}, mental pressure levels \cite{Visser2004}, emotional state \cite{Maehr2005} and may even be able to identify individual users much like a fingerprint \cite{Gamboa2007}. This suggests that cursor tracking data could be a valuable asset when profiling website visitors.\\

This paper presents an Amazon Mechanical Turk (MTurk) study that asked users to complete tasks on live websites using their own hardware in their natural environment. The aim of the study was to explore how cursor tracking data might tell us more about the user than could be measured using traditional means. The study explored several metrics that might be used when carrying out cursor tracking analyses, and used that data to demonstrate that the user's hardware and their browsing intent (search for content \textit{vs.} reading content) could be predicted from cursor movements alone. The study also revealed a number of important practical issues that need to be addressed when attempting to use cursor-tracking data, such as how the user's hardware will affect the movements of the cursor. \\

The structure of this paper is as follows: background work is presented in Section \ref{sec_related_work}, followed by the design of the study and the procedure used in Sections \ref{sec_experiment} and \ref{sec_procedure}. The results of the study are presented in Section \ref{sec_results} and are followed by a discussion of the implications in Section \ref{sec_discussion}. Finally, the findings are summarised and avenues for further research are outlined in Section \ref{sec_conclusion}.


\section{Background}\label{sec_related_work}

One of the most difficult website performance metrics to accurately measure is \textit{user engagement}, generally defined as the amount of attention and time a visitor is willing to spend on a given website and how likely they are to return. Engagement is usually described as a combination of various different characteristics \cite{Attfield2011,Peterson2008,OBrien1999}. Attfield \etal \cite{Attfield2011} discussed several characteristics of user engagement that are difficult to measure including aesthetics and novelty. These would traditionally be measured using physiological sensors (\eg gaze tracking) or surveys. However, it may be possible that this information could be gathered through an analysis of cursor data.\\

Most existing analytic tools focus on a visitor's transition to and from a webpage, using data such as the time spent on the page, the referring page and the page the user went to next. Edmonds \etal \cite{Edmonds2007} argued that ``\textit{within-page activity could inform Web designers about the quality of the content on a page}'', information that is difficult to reliably extract using existing measurements. Edmonds \etal argued for measuring engagement using tools such as cursor-tracking, noting that there is considerable work correlating eye movement and gaze \cite{Chen2001,Huang2012}.\\

Huang \& White \cite{Huang2012} supported this idea and identified several distinct cursor movements in existing work including ``\textit{reading, hesitating, highlighting, marking, and actions such as scrolling and clicking}''. Rodden \etal \cite{Rodden2008} identified several specific behaviours that ``\textit{seemed to indicate active use of the mouse to help the user process the content of the search result page}'', which were following the mouse in the x and y directions and using the mouse to mark the position of interesting results. Rodden's work mostly confirms earlier findings by Claypool \etal \cite{Claypool2001}. However, there was one notable difference: Rodden \etal \cite{Rodden2008} suggested the user would neglect the mouse while reading, whereas Claypool \etal \cite{Claypool2001} suggested that users would frequently use the mouse to follow text when reading. This work suggests that cursor movements, like eye movements, will follow a distinct set of patterns and behaviours that reflect the user's activities.\\

Most existing work is focussed solely on the mouse as a hardware device. However, research has shown that cursor movement properties vary between hardware devices \cite{Accot1999,Hinckley2002}, making it unclear if findings for the mouse can be applied to other input devices. Other factors that have been shown to affect cursor movements include age \cite{Riviere1996}, motor disability \cite{Trewin1999} and mental pressure levels \cite{Visser2004}. These factors may confound the task of interpreting cursor data, but could perhaps be taken into account if known. Unfortunately it must be assumed that nothing is known about the user in the context of online cursor tracking. However, if the user's interactions are observed, it may be possible to predict some of the missing information using statistical models.

\section{Study Design}\label{sec_experiment}

This section presents the design of a study that evaluated cursor tracking metrics `in the wild', taking input hardware into account, in order to identify what they can reveal about the user.

\subsection{Research Questions}\label{sec_research_questions}

The overall research goal was to identify if it is possible to measure user factors (such as interest in the content) solely from cursor data. To explore this, the study was designed to address the following research questions.
\begin{itemize} 
	\item How does the browsing goal of the user (reading \textit{vs.} searching) affect cursor movements?
	\item Does the aesthetic appeal of a website affect cursor behaviour?
	\item Does the user's interest in the content affect cursor behaviour?
	\item Can cursor tracking methods translate across websites?
	\item What is the impact of input hardware on cursor tracking?
	\item Can cursor behaviour be translated into user engagement metrics?
\end{itemize}






\subsection{Design}\label{sec_design}

The study was a mixed-model design with several groups. Two websites were used to ensure that the results would not be limited to a particular context. For each website two interfaces were created: one that would appear as normal and one that was intended to be aesthetically unappealing. Participants were asked to rate their interest in the website's subject matter and were given tasks related to high-interest or low-interest subjects. As two websites were used in the study, there were eight groups in total.

Once allocated to a group, participants were given two different tasks to carry out. These tasks were designed to promote different types of browsing behaviour: \textit{searching} and \textit{reading}. Participants carried out one of each type of task. Therefore, the independent variables of the experiment were: 

\begin{itemize} 
	\item{website;}
	\item{aesthetic appeal of the website;}
	\item{predicted interest in the given task;}
	\item{type of task carried out.}
\end{itemize}

The dependent variables of the experiment were the cursor tracking data (gathered automatically) and engagement data (gathered by survey).\\

The study was designed to promoted a high level of ecological validity. One of the ways this was achieved was by recruiting participants through Amazon Mechanical Turk (MTurk), which allowed participants to carry out the experiment in their natural environment with their own hardware. Another way that ecological validity was promoted was the use of fully interactive websites delivered through a transparent proxy to encourage natural browsing.\\

This potentially introduced several confounding variables, such as gender and hardware that were not controllable. One aim of the study was to find out how such demographic data influenced the cursor data. Therefore demographic data was also gathered. To help ensure a reliable analysis the groups were automatically counter-balanced by gender. Other demographic data were gathered but not controlled.

\subsection{Websites}\label{sec_website_selection}

The websites used in the experiment were Wikipedia (\url{http://en.wikipedia.org}) and BBC News (\url{http://bbc.com}). The websites were chosen for several reasons: 

\begin{itemize}
\item both promote a neutral point of view within their content; 
\item both have sufficiently sized articles and comparable types of content; 
\item both provided search tools for finding content; and 
\item both were large and popular websites at the time of the study (Wikipedia was ranked at \#6 and BBC News at \#81 by Alexa's US traffic rankings). 
\end{itemize}
When selecting content, care was taken with both websites to avoid contentious or provocative subjects which could affect the participant's mood. As Wikipedia's content is user-generated, only protected pages that did not allow editing were used in the experiment.\\


The BBC Website was not the first choice, but as the study was carried out during the 2012 presidential elections, the decision was made to avoid US news websites to avoid giving participants tasks that would conflict with their political beliefs. 
BBC News is generally considered to be a relatively unbiased news website,\footnote{The BBC was famously called a "Stateless Person's Broadcasting Corporation" by a prominent British politician during the Falklands War, who felt the BBC coverage was "unpatriotic".} which was particularly important as biased news could have created a considerable effect on the study which we wanted to avoid.

\subsection{Aesthetics}\label{sec_website_aesthetics}

Two interfaces were created for the websites: one `normal' and one `ugly'. The `normal' version of the website was slightly modified by removing a small number of interface elements which included login forms, user comments, social media links and advertising. External links would appear normal but would no longer function. Figures \ref{fig_wiki_normal} and \ref{fig_bbc_normal} show that the standard interface was retained as much as possible for both sites.\\

In creating the `ugly' versions the aim was to make the sites aesthetically unappealing without drastically affecting the site's usability or layout. This resulted in several changes as follows.

\subsubsection{Font}
{The website fonts were replaced with Comic Sans for text and Impact for Headings. Comic Sans has been shown to be less aesthetically pleasing than other common fonts \cite{Bernard2002}. It has been widely ridiculed as a poor font and there is a well-known movement attempting to raise awareness of its misuse and shortcomings.\footnote{Ban Comic Sans: \url{http://bancomicsans.com}} Impact does not face similar criticisms, but the sharp lines and heavy weight of the font created a contrast with Comic Sans.}

\subsubsection{Colour}
{The colour scheme was changed to create a high contrast. Headings were changed to be light blue, the background dark blue, normal text yellow and hyperlinks red. Yellow is generally considered to be a very poor choice for text as it is difficult to read. However, it should remain readable given the high contrast with the blue background. Style guides generally advocate no more than 2 or 3 colours for a website, so the use of 3 strong colours was expected to produce low aesthetic scores.}

\subsubsection{Advertising}
{A selection of banner advertisements in various standard sizes were downloaded from the internet. These were then injected into pages in appropriate areas such as the sidebar and between sections of text. Around $25\%$ of the banners were animated, resulting in an extremely `busy' appearance.}
	
\subsubsection{Navigation}
{The term `mystery meat navigation' refers to hiding navigation elements until the cursor hovers over them.\footnote{Web Pages That Suck: \url{http://www.webpagesthatsuck.com}} This prevents guests from seeing all options at once, which complicates and slows down navigation. To avoid introducing a serious usability issue, navigation links were coloured to provide minimal contrast with the background, but would change to a high-contrast colour when the cursor was positioned over them.}\\

\begin{figure*}[p]
	\setlength\fboxsep{0pt}
	\setlength\fboxrule{0.4pt}
  	\centering
	\label{fig_all_4_sites}
	\subfloat[Normal Wikipedia website.]{
		\label{fig_wiki_normal}
		\fbox{\includegraphics[width=0.48\textwidth]{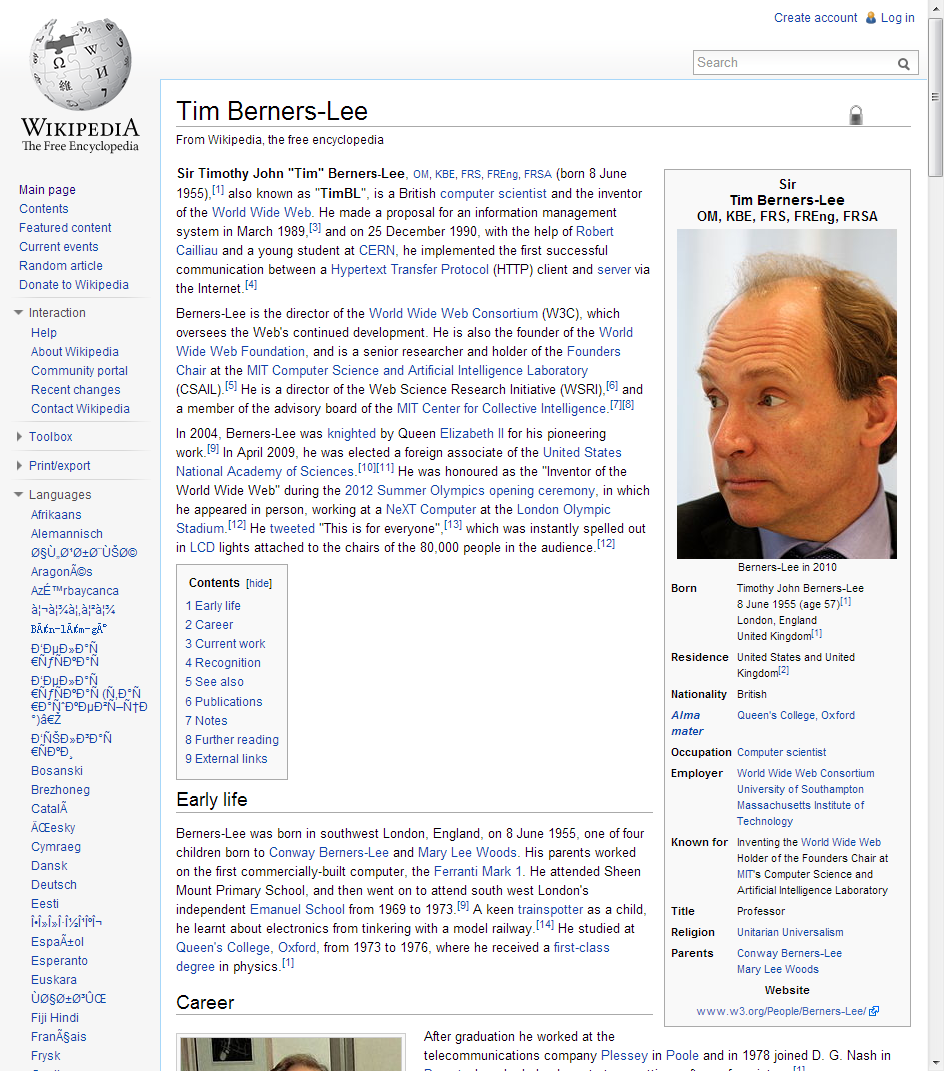}}
	}\hfill
	\subfloat[Modified Wikipedia website.]{
		\label{fig_wiki_bad}
		\fbox{\includegraphics[width=0.48\textwidth]{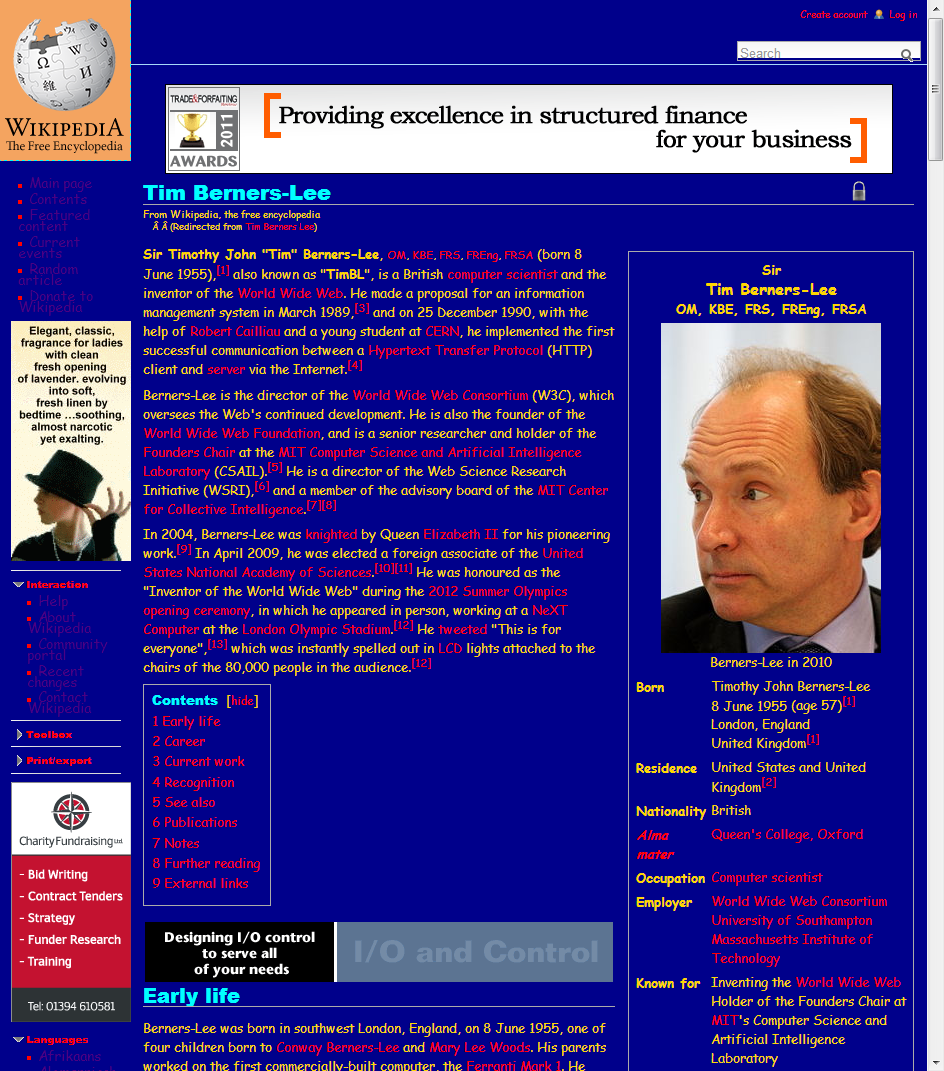}}
	}\\
  	\subfloat[Normal BBC website.]{
		\label{fig_bbc_normal}
		\fbox{\includegraphics[width=0.48\textwidth]{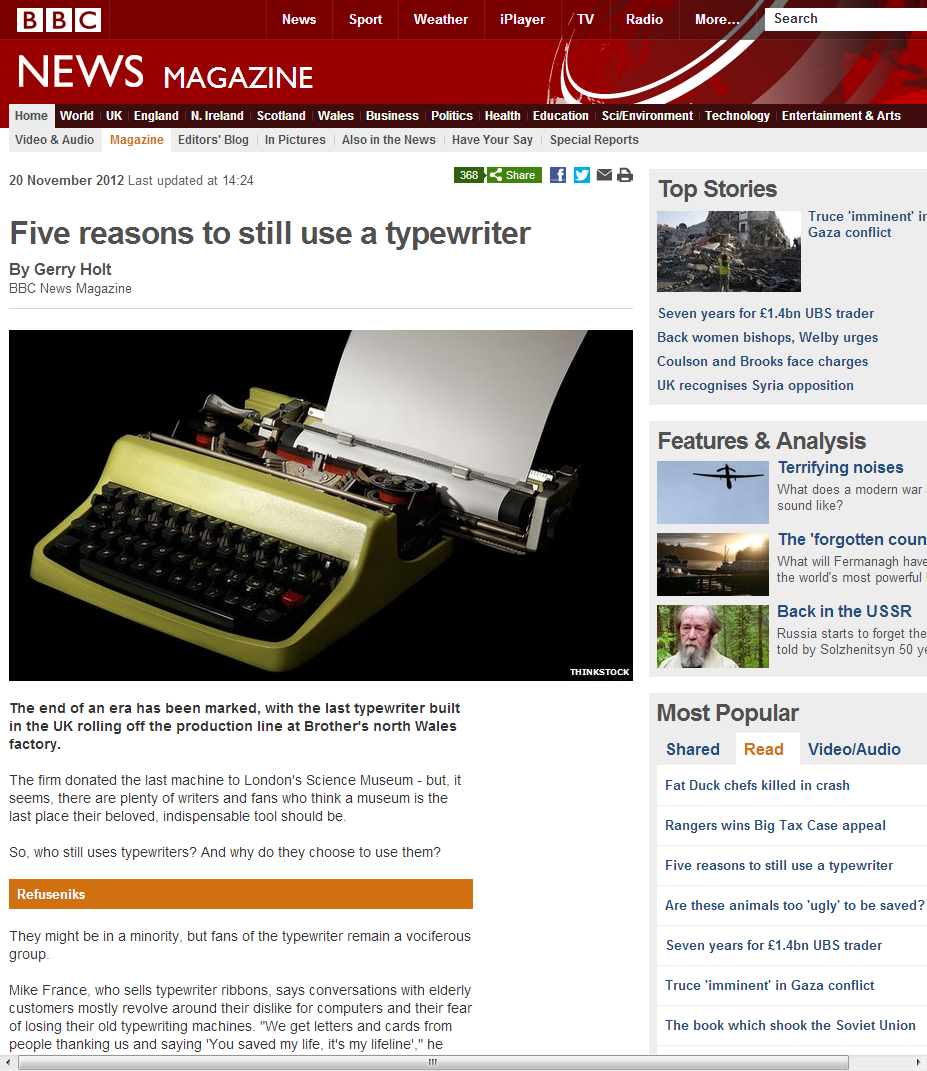}}
	}\hfill
	\subfloat[Modified BBC website.]{
		\label{fig_bbc_bad}
		\fbox{\includegraphics[width=0.48\textwidth]{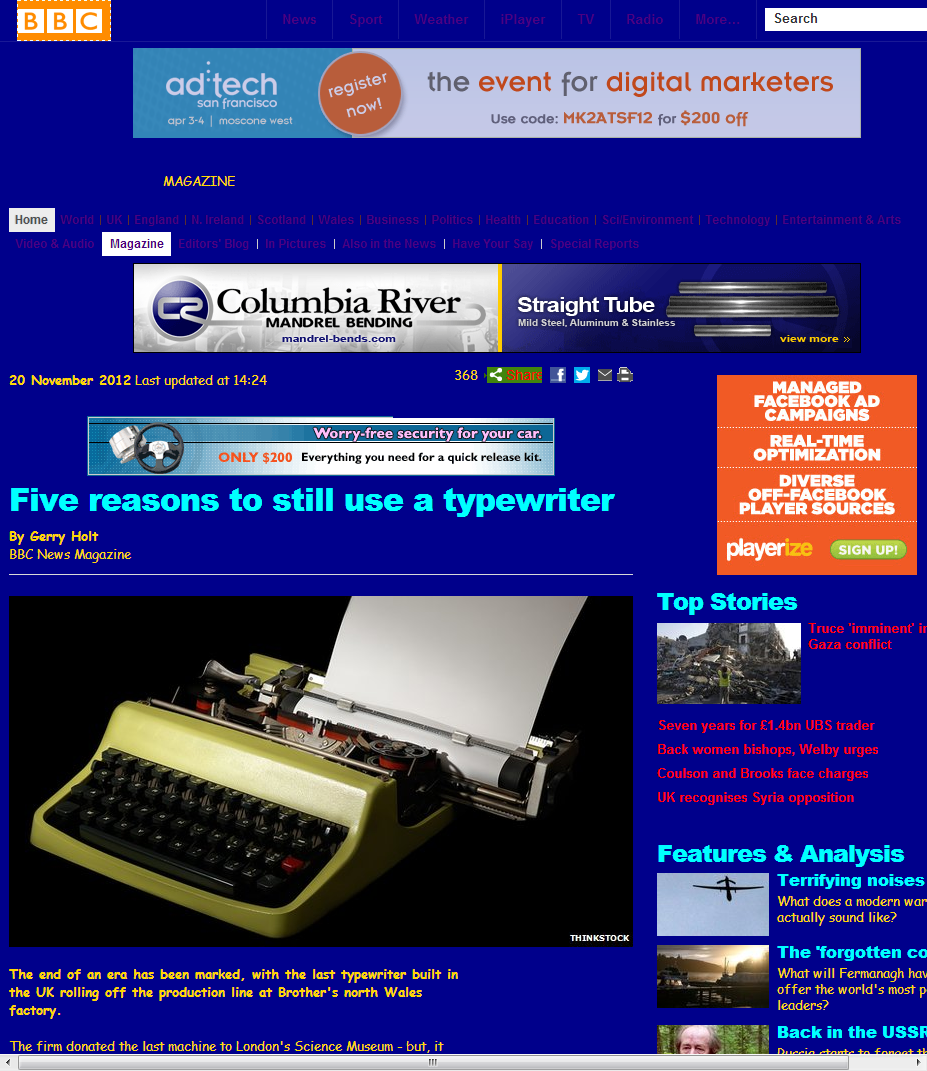}}
	}	
	\caption{Comparison of normal web pages with the `ugly' variants. The latter were modified by changing the colours to create a high contrast, changing the fonts to Comic Sans and Impact, the injection of advertising banners and the obfuscation of navigation elements.}
\end{figure*}

The result of these changes can be seen in Figures \ref{fig_wiki_bad} and \ref{fig_bbc_bad}. While the `ugly' versions of the websites at first appear to be significantly different, the general layout of the pages remained the same. A pilot study was carried out to make sure that the `ugly' websites remained usable but were indeed aesthetically unappealing. Testers reported no significant usability problems and confirmed that the `ugly' variants of the sites were visually unattractive.

\subsection{Tasks}\label{sec_tasks}

Participants were asked to carry out two tasks: a task to promote searching behaviour called the \textit{Search Task} and another to promote close reading called the \textit{Reading Task}. Participants were asked to rate their interest in each website's content categories which allowed participants to be split into `high interest' and `low interest' groups. Wikipedia had 12 content categories while BBC News had 14.

\subsubsection{Search Task}\label{sec_search_task}

The search task involved giving participants a quiz question which could be answered using the website. Questions were selected where answers could be found on both sites. This resulted in 62 questions in total, 61 of which could be answered using Wikipedia and 53 of which could be answered using BBC News. Participants were initially placed on the website's homepage and were expected to use the search feature or navigation to find the answers. Care was taken to ensure that suitable search terms were included when phrasing the questions. Some examples of the questions were:
\begin{itemize} 
	\item What is `Geminoid F'?
	\item In 2008 and 2009 three major car manufacturers withdrew from Formula 1 citing rising costs. Which was the first team to do so in late 2008?
	\item There are many different types of phobia. What is Gephyrophobia a fear of?
\end{itemize}

\subsubsection{Reading Task}\label{sect_quiz_task}

Participants were asked to read an article and write a ``one or two sentence summary of the content in their own words. Some Wikipedia articles were too long for this, so participants were asked to read specific sections. A minimum of two articles were selected for each subject resulting in 24 Wikipedia articles and 28 BBC News articles. Articles were chosen to avoid contentious subjects and based on their length. The average word count of Wikipedia articles was $862$ $(\sigma=260)$ and $840$ $(\sigma=174)$ for BBC News. For example, some of the articles used were:
\begin{itemize}
\item \textbf{Wikipedia} -- The Nile, Section 4.4: The search for the source of the Nile
\item \textbf{Wikipedia} -- Archimedes, Section 1: Biography
\item \textbf{BBC News} -- Olympic torch: Torchbearer proposes during relay
\item \textbf{BBC News} -- Particles point way for Nasa's Voyager
\end{itemize}

\subsection{Measurements}\label{sec_measurements}

Three core sets of measurements were taken during the study: cursor, engagement and demographic.

\subsubsection{Cursor Metrics}\label{sec_cursor_metrics}

Rodden \etal \cite{Rodden2008} identified 3 ways in which users could use the cursor when reading web search results:

\begin{itemize}
\item The mouse could follow the eye horizontally;
\item The mouse could follow the eye vertically;
\item The mouse could `mark' results.
\end{itemize}

There is a range of things the cursor could do while the user reads the page. The mouse following the eye horizontally would likely be signified by a slow and steady horizontal movement in the direction of the text followed by a `flick' back to the next line. Following the text vertically would result in a slow mouse movement in a downwards direction. Marking the text, or interesting areas, would perhaps be signified by long periods of inactivity in whitespace. Other behaviours might also be expected \cite{Claypool2001,Huang2011}, but as most existing work is focussed on search engines, new behaviours could be observed in the reading task. Existing work shows that the most interesting areas of observation for cursor tracking are the speed and frequency of movements, the number of mouse clicks and the amount of time that the cursor lies still. This led to the creation of the following 5 measures:
\begin{itemize}
\item {\bf Movement Speed (\vmove)}: {Average speed over all movements in pixels per second.}

\item {\bf Movement Rate (\rmove)}: 
{Number of distinct movements made per second.}

\item {\bf Click Rate (\rclick)}: 
{Number of mouse clicks that were made per second.}

\item {\bf Pause Length (\tstill)}: 
{Average length of a pause in seconds.}

\item {\bf Percentage of Time Still (\pstill)}: 
{Percentage of time where the cursor did not move.}

\end{itemize}

Cursor data was gathered as a system of \textit{x} and \textit{y} coordinates gathered at a rate of 24fps as suggested by Leiva \cite{Leiva2007}. The high recording rate allowed even very small pauses to be identified, which allowed the data to be split into distinct movements. The movements were then categorised as either pauses, movements or scrolls. A pause happened when the cursor was stationary and a movement occurred when the cursor moved. If $99\%$ or more of the movement was vertical then that movement was reclassified as a scroll. As the websites had very different vertical sizes, and as the advertisements injected into the `ugly' websites increased their vertical size, vertical scroll movements would be likely to confound the results and as such were not included in the analysis.

\subsubsection{Engagement}


Attfield \etal \cite{Attfield2011} defined engagement as \textit{``the emotional, cognitive and behavioural connection that exists, at any point in time and possibly over time, between a user and a resource''}. O'Brien \& Toms \cite{O'Brien2008} defined engagement as \textit{``characterized by attributes of challenge, positive affect, endurability, aesthetic and sensory appeal, attention, feedback, variety/novelty, interactivity, and perceived user control.''}. Engagement is a desirable property for websites, as engaged users are \textit{(1)} likely to view more content and spend more time on a website and \textit{(2)} more likely to return to the website at a later date. Engagement is difficult to measure however, and is normally assessed using eye tracking methods. Studies have shown that gaze and cursor often correlate, so it may be possible to determine engagement from cursor movements.\\

O'Brien \& Toms \cite{O'Brien2010} created a survey to measure user engagement, defining it in terms of attention, usability, aesthetics, endurability and novelty. Attfield \etal \cite{Attfield2011} built on this work and argued that the relevant factors were attention, affect, aesthetics, endurability, novelty, control, reputation and user context. For some of these factors there are well-known surveys that can be used for assessment, such as the PANAS scale for measuring affect \cite{Watson1988}.\\

In this study engagement was measured using two tools. Firstly, the PANAS \cite{Watson1988} was used to measure positive and negative affect. Secondly focussed attention, perceived usability, novelty, aesthetics and involvement were measured based on the work of O`Brien \& Toms \cite{O'Brien2010}. The wording of their survey components was modified to suit the study by removing the shopping-task context.

\subsubsection{Demographic Data}\label{sec_demographic_measurements}

\begin{figure}[t]
	\begin{center}
		\leavevmode
		\includegraphics[width=1\columnwidth]{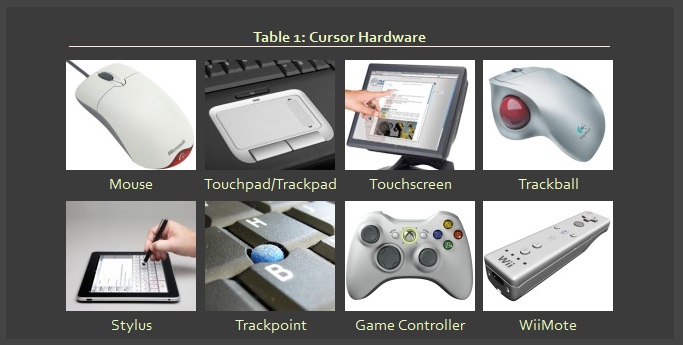}
		\caption{The input hardware question as presented in the study.}
		\label{fig_hardware_demographic}
	\end{center}
\end{figure}

Demographic data was gathered to identify any interesting effects. All participants were required to be US residents to participate in the study. Participants were asked to provide their age, gender, handedness, computer experience, media consumption habits and input hardware. To help participants identify their input hardware, pictures of various devices were provided, as shown in Figure~\ref{fig_hardware_demographic}.

\subsection{Implementation}\label{sec_implementation}

Implementing the study required two components: a proxy server to manipulate webpages and a cursor tracking system. The cursor tracking software used was the open-source system Simple Mouse Tracking 2 (\smt), created by Leiva \cite{Leiva2007}. \\

The proxy was implemented in PHP and was transparent to the user (\ie participants did not have to change their browser settings or install software). The proxy would take a standard HTTP GET request containing the address of the target website and any additional parameters to include in the query. We called this proxy a \textit{proxy trap} as one of its primary functions was to ensure that visitors could not leave the original site by clicking links. When the proxy was called it carried out the following actions:

\begin{enumerate} 
	\item Reconstruct the target URL from the HTTP request.
	\item Fetch the target page using PHP cURL.
	\item Strip out any `banned' elements, such as html {\ttfamily{<div>}} elements containing social media links or javascripts which loaded dynamic content.
	\item Remove the target of anchors which led to external sites and redirect internal links to point back to the proxy (preventing users from leaving the proxy).
 	\item Append an overriding stylesheet to create the normal/ugly interfaces.
	\item Insert javascript that enabled the \smt tool.
	\item Insert an HTML element at the top of the webpage which was used to send messages or links to the participant regarding the study.
	\item Send the resulting page to the participant's browser.
\end{enumerate}

The advantage of this system was that only the text of a given website would pass through the proxy, leaving images and scripts to load from the original source. This resulted in good performance with no noticeable lag when comparing the proxy to the original website.




\section{Procedure}\label{sec_procedure}

Amazon Mechanical Turk (MTurk) was used to carry out the experiment. The study was split into two parts, with Wikipedia carried out first $(n=160)$ and BBC News second $(n=165)$. Work units were placed on MTurk with a unique code. MTurk workers would read the description of the study, accept the work unit, then click a link to the study website where they would input their unique code. Workers were informed that they would not be paid if they participated in the study multiple times.\\

At the start of the study participants were asked to verify their consent with regard to the data that would be gathered. Following this the participant would be asked their gender, which was used to balance the gender distributions in each group\footnote{On average, each Wikipedia group had 18 males and 22 females and each BBC group had 22 males and 19 females.}. The 4 groups varied in terms of task interest and website aesthetics. Participants were then asked to fill out a survey based on their interests relative to the website's content. Depending on their group, they were then given high-interest or low-interest content.\\

Before the first task the participant was asked to fill out a PANAS questionnaire to assess their affect prior to the first task. The two tasks (search and reading) were delivered in a random order. Both followed the same format, which started by describing the task then directing the participant to the proxy. A banner at the top of the website would provide a reminder of the task and included a button to press when the participant was finished. When pressed the button would take the participant to a page where they could type in their answer. For the reading task a word counter was provided that would limit the amount participants could write by turning green after 10 words and red after 50 words. With each task completed, participants were administered another PANAS followed by the engagement survey. The questions in both surveys were randomly ordered for every participant. At the end of each task participants were given the option to provide comments or feedback if they wished. Finally, participants were asked to fill out the demographic survey.\\

With the work completed, participants were then given a second code to insert back into MTurk to complete the work unit. This process took around 15-25 minutes per participant and each participant was paid \$2.50. The study was split up into batches of 80 participants which required around 2 hours each to complete.

\subsection{Participant Validation}\label{sec_participant_validation}

To verify that participants had correctly carried out their work two methods were employed. Firstly, the response codes entered by participants to complete the work were compared to the codes originally provided to them. Any codes which did not match suggested that the participant had simply entered a random code to complete the work without taking the study. Only 1 participant did this and their work unit was simply rejected and returned to MTurk for another worker to complete.\\

The second stage of verification was to manually check participant's answers to both the search and reading tasks. This was labour-intensive but revealed that the majority of the MTurk workers completed their tasks to a high standard. Of 325 workers, only 1 participant's work was considered so poor that it had to be rejected. Although it was not required, a large number of participants left comments on their experience.\\

The IP addresses of workers were also checked which revealed two duplicates. While this could be people with multiple accounts, it could also show people sharing an internet connection and was not considered suspicious.

\section{Results}\label{sec_results}

The study examined the effects of four dependent variables (task, website, predicted interest and interface version) on two sets of independent variables (cursor movement and engagement).

\subsection{Cursor Movements}\label{sec_cursor_movements}

The five cursor metrics measured were analysed in turn using a multi-factorial ANOVA where the within-subjects factor was task and the between-groups variables were website, interface version and predicted interest. Participant input hardware was included as a covariant. Note that the results showing the effects of hardware on the model are presented later in this section.

\subsubsection{Movement Speed}
{No effects were found of task, website, interface version or predicted interest on \vmove and no interactions were found.}

\subsubsection{Movement Rate}
{Task was found to have a significant affect on \rmove $(F(1,244)=18.24$, $p<.001$, $\omega^2=.07)$ but no other effects or interactions were found.}

\subsubsection{Click Rate}
{No effects were found of task, website, interface version or predicted interest on \rclick and no interactions were found.}

\subsubsection{Pause Length}
{Website was found to have an effect on the \tstill $(F(1,244)=3.97$, $p<.05$, $\omega^2=.02)$, while no effects were found for task, predicted interest or interface. Task type and predicted affinity interacted to affect \tstill $(F(1,244)=3.97$, $p<.05$, $\omega^2=.02)$. No other interactions were found.}

\subsubsection{Percentage of Time Still}
{Task was found to affect \pstill $(F(1,244)=29.57, p<.001, \omega^2=.11)$, but no effects were found for website, interface or predicted interest and no interaction effects were found.}\\

The results suggest that the two interfaces made little difference to the cursor movements. Predicted interest seemed to have no effect at all, but interacted with task type to affect \tstill. Website only had an effect on \tstill. The results showed that the type of task affected both \rmove and \pstill, but was not found to affect the other cursor metrics.\\

The ability to distinguish between searching and reading behaviour could be very useful in understanding website visitors. Logistic regression\footnote{For information on logistic regression see Peng \etal \cite{Peng2002}.} can be used to evaluate if the user's activity can be predicted from their cursor movements. The variables with the strongest effects were \rmove and \pstill. However, the model produced by these variables was not found to be a good fit for the data. Of the remaining three variables, only \tstill was found to correlate to task type $(r(550)=-.33$, $p<.001)$. The collinearity of \rmove, \pstill and \tstill was found to be acceptable ($\overline{VIF}$=$3.22$) and so all three variables and their interactions were used to carry out a binary logistic regression with the backwards-stepwise LH method. The resulting model is shown in Table \ref{tab_task_type_regression} and was able to correctly identify user activity in $74.5\%$ of cases.

\begin{table}[tb]
	\centering
	\begin{threeparttable}[c]
		\caption{Binary Logistic Regression Analysis of Task using \rmove, \tstill and \pstill.}
		\label{tab_task_type_regression}
		\small
		\begin{tabularx}{\columnwidth}{X r@{.}l r r r r} 
			\toprule
								& $\beta$ & 					& $SE \beta$ 	& $W$& $e^\beta$\\
			\midrule			
			Constant 			& 11&59\superscript{***} 		& 2.89	 		& 16.13				& NA \\ 
			\rmove 				& -10&33\superscript{***} 		& 2.15	 		& 23.07 			& 0.00 \\
			\tstill 			& -7&93\superscript{**} 		& 3.00	 		& 7.00	 			& 0.00 \\
			\pstill 			& -13&49\superscript{***} 		& 3.08	 		& 19.24 			& 0.00 \\
			\tstill*\pstill 	& 7&71\superscript{*} 			& 2.99	 		& 6.65	 			& 2238.83 \\
			\rmove*\pstill 	& 13&90\superscript{***} 		& 2.72	 		& 26.07 			& 1082822.66 \\
			\tstill*\rmove*\pstill & 3&89\superscript{**} 	& 1.29	 		& 9.12	 			& 49.00\\
			\bottomrule
		\end{tabularx}
		\begin{tablenotes}[flushleft]
			\item \footnotesize Note: Model produced after 2 iterations, 1 interaction removed. Model: $\chi^2(6)=186.82$, $p<.001$. Cox \& Snell $R^2$=$.29$, Nagelkerke $R^2$=$.38$. Hosmer-Lemeshow goodness-of-fit test: $(\chi^2(8)=9.04$, $p=.34)$. Sig: {[*]}=$p<.05$, {[**]}=$p<.01$, {[***]=$p<.001$}. {[}$e^\beta${]}=Odds Ratios, {[}$W${]}=Wald's $\chi^2$.
		\end{tablenotes}
	\end{threeparttable}
\end{table}

\subsection{Engagement Data}\label{sec_engagement}

To examine the effects of the four factors on the engagement data each engagement component was taken in turn and analysed with the multi-factorial ANOVA.

\subsubsection{Affect}
{No main effects were found from any of the factors on positive or negative affect. However, an interaction effect was found of task type and predicted interest on positive affect $(F(1,244)=4.35, p<.05, \omega^2=.02)$. The data suggest that positive affect increased during the reading task when participants were given subjects that match their predicted interest levels.} 

\subsubsection{Focussed Attention}
{An interaction effect between predicted affinity, interface version and task type was found $(F(1,244)=4.03, p<.05, \omega^2=.02)$, but no other effects or interactions were found.}

\subsubsection{Perceived Usability}
{Task type affected perceived usability $(F(1,244)=5.15$, $p<.05$, $\omega^2=.02)$. Interface version, predicted interest and website were not found to have an effect. Task type and interface version created an interaction effect on perceived usability $(F(1,244)=4.37$, $p<.05$, $\omega^2=.02)$. Task type, interface version and website interacted to affect perceived usability $(F(1,244)=5.97$, $p<.05$, $\omega^2=.02)$.}

\subsubsection{Aesthetics}
{Interface version (normal \textit{vs.} ugly) was \textit{not} found to affect the aesthetic scores, but website was found to have an effect $(F(1,244)=20.47$, $p<.001$, $\omega^2=.08)$. Task type and predicted interest were not found to affect aesthetic scores, but together created an interaction effect $(F(1,244)=4.6$, $p<.05$, $\omega^2=.02)$. No other interactions were found.}

\subsubsection{Novelty}
{None of the factors were found to affect novelty.}

\subsubsection{Involvement}
{None of the factors were found to affect involvement.}\\

There were two surprising results. Firstly, the two interface versions were not found to have significantly different aesthetics scores, yet the two websites were with a surprising effect size $(\omega^2=.08)$. Second, the participant's predicted interest did not appear to influence their effect ratings, which would be expected if they were actually given more or less interesting tasks. However some effect was found when considering the reading task.

\subsection{Demographic}\label{sec_demographic}

The data was checked for correlations between the gathered demographic data and various cursor movement data as described in Section \ref{sec_measurements}. All correlations were carried out over the whole data set controlling for task, website, interface version and task ordering. The interface version must be included as a control variable as the addition of advertisements on the `ugly' webpages resulted in more vertical space and therefore more vertical scrolling.\\

There were considerable differences in speed between participants. While we checked computer experience, the speed of participants did not correlate strongly to this; interestingly, speed correlated strongly to the age of the participant instead.\\

To address this the movements were divided by the time, giving us time-independent data. The data was then divided by the total number of movements to provide the percentage of movements in a minute, accounting for the differences in activity rate between participants. The data represents the percentage of each type of cursor movement a participant made over an average minute.

\subsection{Predicting Engagement}\label{sec_predicting_engagement}

Cursor data were checked for correlations with the engagement data, but no correlations were found. The tests were re-run to control for any variance caused by cursor hardware, task and website; however there still appeared to be no correlation between the cursor metrics tested and the subjective survey responses of the participants.

\subsection{Hardware}\label{sec_hardware}

\begin{table}[tb]
	\centering
	\caption{Participant cursor hardware.}
	\small
	\label{tab_hardware}
	\begin{tabular}{l r r} 
		\toprule
		& Wikipedia & BBC News\\
		\midrule
		Mouse 		& 110 	& 94 \\
		Trackpad 	&  45 	& 68 \\
		Touchscreen &  4 	&  3 \\
		Trackball   &  1 	&  0 \\
		\bottomrule
	\end{tabular}
	\vspace{-1em}
\end{table}

Table \ref{tab_hardware} shows that most participants used either a mouse or trackpad, and due to the small amount of data for other hardware types, the analysis was carried out using only mouse and trackpad data. Included as a co-variate in the model described in the `Cursor Movements' section, hardware was found to affect:

\subsubsection{Movement Speed}
{Input hardware was found to affect \vmove $(F(1,244)=39.19$, $p<.001$, $\omega^2=.14)$.}

\subsubsection{Movement Rate}
{Input hardware was found to affect \rmove $(F(1,244)=26.61$, $p<.001$, $\omega^2=.10)$.}

\subsubsection{Click Rate}
{Input hardware was found to affect \rclick $(F(1,244)=47.53$, $p<.001$, $\omega^2=.16)$.}

\subsubsection{Pause Length}
{Input hardware was found to affect \tstill $(F(1,244)=17.75$, $p<.001$, $\omega^2=.07)$. Task was found to interact with hardware to affect \tstill $(F(1,244)=11.52$, $p<.01$, $\omega^2=.05)$.}

\subsubsection{Percentage of Time Still}
{Input hardware was found to affect \pstill $(F(1,244)=23.39$, $p<.001$, $\omega^2=.09)$.}\\

These results show that input hardware has a strong effect on the behaviour of the cursor. Additionally, task and hardware interacted to affect \tstill. It would be useful if input hardware could be determined from cursor movements. This can be tested with binary logistic regression. Both \tstill and \pstill were found to correlate quite strongly with \rmove, so both could not be included in the regression. The results showed that \vmove, \rmove and \rclick had the strongest effect and a low degree of collinearity $(\overline{VIF}=1.07)$, so they were used to carry out a binary logistic regression using the backwards stepwise LH method. Table \ref{tab_hardware_regression} shows the resulting model which is able to correctly identify input hardware in $75.5\%$ of cases.

\begin{table}[tb]
	\centering
	\begin{threeparttable}[c]
		\caption{Binary Logistic Regression  Analysis of Input Hardware using \vmove, \rmove and \rclick.}
		\small
		\label{tab_hardware_regression}
		\begin{tabularx}{\columnwidth}{X r@{.}l r r r r} 
			\toprule
								& $\beta$ & & $SE \beta$ & $W$ & $e^\beta$\\
			\midrule
			Constant			& -1	& 255\superscript{***}	& 0.268	& 21.86		& NA\\
			\rmove 				& -1	& 758\superscript{***}	& 0.446	& 15.57		& 0.17 \\
			\vmove 				&  0	& 001\superscript{***}	& 0.000	& 17.69		& 1.00 \\
			\rmove * \rclick    &  0	& 003\superscript{**}	& 0.001	& 10.57		& 1.00 \\
			\vmove * \rclick	&  -1	& 255\superscript{***}	& 0.000	& 8.21		& 1.00 \\
			\bottomrule
		\end{tabularx}
		\begin{tablenotes}[flushleft]
			\item \footnotesize Note: Model produced after 4 iterations, 1 variable and 2 interactions removed. Model: $\chi^2(4)=133.67$, $p<.001$. Cox \& Snell $R^2$=$.22$, Nagelkerke $R^2$=$.30$. Hosmer-Lemeshow goodness-of-fit test: $(\chi^2(8)=8.25$, $p=.41)$. For $p$ values, {[**]}=$p<.01$, {[***]=$p<.001$}. {[}$e^\beta${]}=Odds Ratios, {[}$W${]}=Wald's $\chi^2$.
		\end{tablenotes}
	\end{threeparttable}
\end{table}


\begin{figure*}[t]
	\begin{center}
		\leavevmode
		\includegraphics[width=1\textwidth]{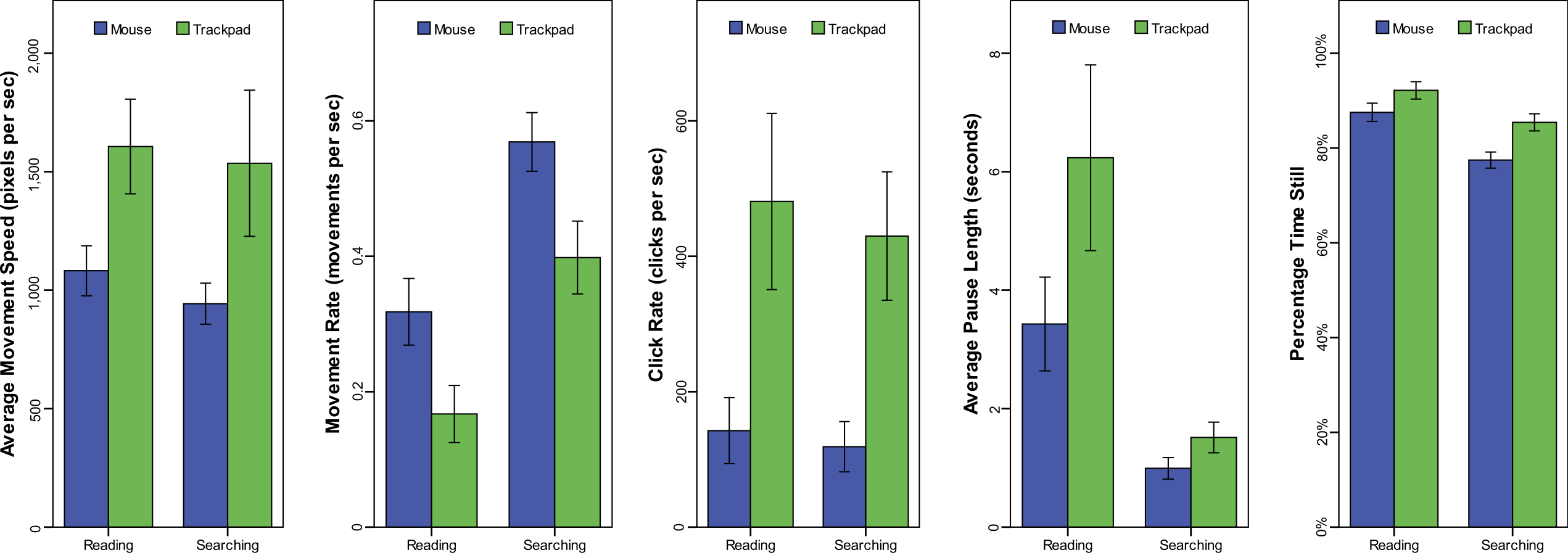}
		\caption{Graphs showing the effects of input hardware and task on the cursor metrics \vmove, \rmove, \rclick, \tstill and \pstill.}
		\label{fig_profiling_data}
	\end{center}
	\vspace{-1em}
\end{figure*}

\section{Discussion}\label{sec_discussion}

There were two important findings from this study. Firstly, the study revealed some of the differences in cursor movements between users searching and users reading. Secondly, the study identified some of the differences between the cursor behaviours of a mouse and a trackpad. The regressions provided in this paper show that both the user's activity (reading or searching) and input hardware (mouse or trackpad) can be predicted from the behaviour of the cursor. This could have several interesting applications in a range of domains.\\

An obvious application is to form a better understanding of website guests. The ability to differentiate between a user searching and reading could have benefits in several areas, such as when carrying out `in the wild' usability studies. For example, if using cursor data to evaluate the interface using the methods described by \cite{Arroyo2006a}, tracking data could be split into `searching' and `reading' to provide a better understanding of the user's goals, allowing usability issues that affect one type of user to be isolated.\\

This information could also be useful as a part of content management; Wikipedia for example would be expected to have a large number of guests who are both searching and reading. Wikipedia promotes certain articles on its front page, but given two pages with equal traffic, which one should be promoted? If the traffic to both pages could be split into readers and searchers then the page with the most readers would likely be the page to promote. This could also be used to promote `searched' pages above `read' pages on the site's search engine. Another example is that shopping sites could direct advertising towards products based on other products that guests took time to read, instead of assuming that all previously-viewed products were equally indicative of the type of products that the customer wants to buy.\\

Identifying the input hardware of users is an important step in attempting to interpret cursor data as the results of this work have clearly shown that input hardware has an effect on other measurements. The ability to identify cursor hardware will be highly important to making reliable observations from cursor data. Assuming that the findings presented here can be extended to detect other types of hardware, in particular touch screens or stylus interactions, then it could also have interesting applications in website design. While the results of this study suggest that the mouse and trackpad are the most common input devices, touchscreen technology is becoming increasingly common. Touchscreen technology renders the traditional cursor somewhat obsolete: the finger becomes the cursor, and as such common website design elements such as drop-down menus and hover-to-enlarge images might not be suitable ways to interact. The ability to identify the user's hardware could be useful in the same way that developers would previously monitor browser versions; although in this case it would not be to provide workarounds for browser-specific bugs, but to provide a slight variant of the website that is optimised for the user's hardware to provide more natural interactions.\\

The most surprising result from the study was that the `ugly' variants of the sites did not result in lower aesthetics scores, yet the BBC News website was considered to be more aesthetically pleasing than Wikipedia. This finding is made even more unusual when participant comments are taken into consideration:
\begin{itemize} 
	\item {[}Wikipedia{]} ``The website was simply awful. Ads flashing everywhere, poor text colors on a dark blue background.''
	\item {[}Wikipedia{]} ``The webpage was entirely blue. I don't know if it was supposed to be like that, but it definitely detracted from the browsing experience.''
	\item {[}Wikipedia{]} ``Only that it has slightly interested me as to why the site made it intentionally difficult.''
	\item {[}BBC News{]} ``The font and the colors used were really distracting to me.''
	\item {[}BBC News{]}``Multiple font colors within the same page = hard to read.''
	\item {[}BBC News{]} ``The website's layout and colour scheme were a bitch to navigate and read.''
	\item {[}BBC News{]} ``Comic sans is a horrible font.''.
\end{itemize}

\begin{table}[tb]
	\centering
	\caption{Summary of comments left by participants regarding the interface. Total includes comments that did not mention the interface.}
	\small
	\label{tab_comments}
	\begin{tabularx}{\columnwidth}{X r r r r}
		\toprule
		Comments & \multicolumn{2}{c}{Wikipedia} & \multicolumn{2}{c}{BBC News} \\
		\cmidrule(r){2-5}
		& Normal & Ugly & Normal & Ugly \\
		
		\midrule
		Positive 	&  0 &  1 &  0 &  0 \\
		Neutral 	&  1 &  0 &  0 &  0 \\
		Negative 	&  1 & 30 &  0 & 23 \\
		Total 		& 61 & 70 & 44 & 51 \\
		\bottomrule
	\end{tabularx}
	\vspace{-1em}
\end{table}

Participants were not prompted to discuss the interface in any way; the comment box was simply titled ``Comments (optional)''. Yet as Table \ref{tab_comments} shows, a large number of participants left negative comments about the `ugly' versions of the interface. While a single person ``loved the colours'', this may have been sarcasm. These comments suggest that the aesthetics scores do not accurately reflect the aesthetics of the interfaces. Further work would be needed to fully explore this unexpected result.\\

There was no correlation between the predicted interest in the task and a participant's reported interest in the task. There were also no significant correlations between any of the cursor metrics and the engagement data from the surveys. While these results are not suspect, the lack of correlation between aesthetic scores and user comments creates questions about the reliability of the engagement survey data. Due to the between-groups design of the experiment, participants had no basis of comparison for the majority of the engagement metrics, which could have impacted on their answers. Another possible source for interference is the Hawthorne effect; participants were notified that their cursor was being tracked, which could have influenced both their cursor movements and the engagement measures. While it is possible that MTurk users might have `randomly' clicked through the survey, we do not believe this was the case; the manual result verification showed that all questions were answered to a high standard, and the majority of participants $(70\%)$ took extra time to answer optional parts of the survey. Therefore, we believe that the mouse tracking data is sound but there is a possibility that the engagement data is not reliable. Further studies would be needed to explore this finding.\\

If the engagement data \textit{is} reliable, then the results suggest there is no relationship between these cursor and engagement measures. Yet these measures represent only a fraction of the data that can be extracted from the cursor; alternative metrics are likely to provide even more information. To further investigate ways to predict engagement from cursor data, one approach might be to apply statistical clustering methods to cursor data streams. This could reveal specific types of movement in the same way that gaze tracking studies observe saccades and fixations. 
Recent work by Arapakis \etal \cite{DBLP:conf/cikm/ArapakisLV14} suggests this is the case, as they were able to correlate certain activities with subjective engagement metrics, specifically focus attention and affect.\\

There could be several reasons for these negative results (beyond the typical drawbacks associated with using questionnaires as discussed above): flawed methodology, a non-existent signal or use of the wrong measures. In terms of the methodology, due to between-groups design of the experiment participants had no basis of comparison for the engagement metrics, which could have impacted their answers. However, performing a within-subjects study where participants interacted with two variants of the website could have potentially introduced confounding variables. It could also be that the experience induced by the `ugly' version of the interface was not negative enough for users to become annoyed, and thus down-marking it. However experimenting with an uglier interface would not be valid (unless the aim is to test this hypothesis) as measuring user engagement only makes sense if the experience is in principle positive. Although usability is an important characteristic of user engagement, usability issues must be fixed first before thinking about user engagement measurements.\\

In hindsight, this study demonstrates that designing experiments to obtain reliable insights about user engagement and its measurement remains highly challenging. Finally, not finding a signal may simply mean that some of the metrics used were not the correct ones. This is in fact what we believe is happening here. The cursor metrics were not \emph{the right} ones to differentiate between the levels of engagement experience, which is supported by recent work by Arapakis \etal \cite{DBLP:conf/cikm/ArapakisLV14} which showed that more complex mouse metrics, based on mouse gestures, \emph{did} correlate with focus attention and affect engagement metrics.\\

\section{Conclusion}\label{sec_conclusion}

The results of this study revealed how a user's activity (searching for information or reading) and their cursor hardware (mouse or trackpad) can be predicted purely from cursor tracking data. While no conclusive evidence was found linking participant engagement to cursor data, there are still several interesting paths for future work such as evaluating further cursor data metrics and linking these results with traditional web analytic data. The 5 simple measurements evaluated in this study have allowed detection of user task and input hardware; yet it is expected that there will be a much greater range of cursor metrics not yet evaluated which can provide a deeper understand of the user (as shown in recent work~\cite{DBLP:conf/cikm/ArapakisLV14} in the context of user engagement self-report measures~\cite{bookUE}). Future work should focus on exploring other metrics. Another future direction of this work would be to combine it with existing website analytics such as `page views' and `time spent on a page' to improve the reliability of these models. More work should be carried out to evaluate if the input hardware identification model presented here can be extended to other types of input hardware such as touchscreens.\\

This work adds to a growing body of research that suggests that tracking cursor movements might lead to new techniques that would allow powerful and inexpensive user profiling and usability testing. It is hoped that this work will form the basis of a better understanding of how we use the cursor to interact, and how this information can be used to improve website designs and services.


\section{Acknowledgements}

\anon[** Removed for review copy. **]{We would like to thank Diego Sáez Trumper for his assistance running this study, Luis A. Leiva for his work on \smt (\url{http://code.google.com/p/smt2/}) and his input, and our patient  Mechanical Turk participants.}

%
%
%
%
%


\newpage

\end{document}